\documentclass{cernrep}
\usepackage{graphicx,here}
 
\begin{document}

\title{Questions on Uncertainties in Parton Distributions}
 
\author{R. S. Thorne$^a$, H. B\"ottcher$^b$, A. M. Cooper-Sarkar$^c$, 
B. Reisert$^d$, V. Shekelyan$^d$, W. J. Stirling$^e$, D. R. Stump$^f$, 
A. Vogt$^g$}
 
\institute{${}^a$Cavendish Laboratory, University of Cambridge, Cambridge, UK\\
${}^b$DESY, Platanenallee 6, D-15738, Zeuthen, Germany\\
${}^c$Department of Physics, University of Oxford, Oxford, UK\\
${}^d$Max-Planck-Institut f\"ur Physik, Munich, Germany\\
${}^e$Department of Physics and Institute of Particle Physics Phenomenology, 
University of Durham, Durham, UK\\
${}^f$Department of Physics and Astronomy, Michigan State University, 
East Lansing, MI 48824, USA\\
${}^g$Nikhef Theory Group, Kraislaan 409, 1098 SJ Amsterdam, The Netherlands}

%-----------------------------------------------------------------------
% If your printer does not reproduce dimensions exactly, it may be
% necessary to remove the % signs and adjust the dimensions in the
% following commands:
%
%     \setlength{\textheight}{24cm}
%     \setlength{\textwidth}{16cm}
%
% Similarly for the following, if you need to adjust the positioning
% on the paper:
%
%         \setlength{\topmargin}{-1cm}
%         \setlength{\oddsidemargin}{0pt}
%         \setlength{\evensidemargin}{0pt}
%------------------------------------------------------------------------
 
\def\GeV{{\rm GeV}}
 
\maketitle % this produces the title block
 
\begin{abstract}
A discussion is presented of the manner in which uncertainties in parton 
distributions and related quantities are determined. One of the central 
problems is the criteria used to judge what variation of the
parameters describing a set of partons is acceptable
within the context of a global fit. Various ways of addressing this 
question are outlined.  
\end{abstract}
 
\section{Introduction}

The procedure of determining parton distributions by so-called global fits to 
data, mainly structure functions, is long established 
\cite{GRV}-\cite{ref:giele}. However, it is a 
rather more recent development to try to determine the errors on 
these distributions at the same time. This has come about for a number of 
reasons. Firstly, the sheer amount of data (full references in \cite{stat})
sensitive to various parton 
distributions, and the precision of this data, has become such that an 
accurate determination of all parton distributions is possible (with some 
problems only in difficult to reach regions of phase space, e.g. $x$ very 
near to 1). Secondly, the understanding of the experimental errors on this 
data has reached a new level of sophistication, with the systematic errors 
being understood far better in terms of their separate sources and 
correlations. Lastly, the theoretical understanding at NLO in $\alpha_S$ has
improved so that subtleties due to e.g. heavy quarks are now understood. 

There are many issues in the determination of errors on parton 
distributions, and a discussion of these may be found in \cite{stat}.          
However, one of the main outstanding problems, and the focus of a 
discussion session at this meeting, is the manner in which one 
determines precisely the size of the errors. 

\section{Quality of fit}

The quality of the fit to a set of data 
is generally presented in terms of the $\chi^2$. Taking fully into account the 
correlated systematic errors this may be calculated by the covariance 
matrix method, i.e. the covariance matrix is constructed as
\begin{equation}
C_{ij} = \delta_{ij} \sigma_{i,stat}^2 + \sum_{k=1}^n \rho^k_{ij}
\sigma_{k,i}\sigma_{k,j},
\label{k}
\end{equation}
where $k$ runs over each source of correlated systematic error
and $\rho^k_{ij}$ are the correlation coefficients. The $\chi^2$ 
is then defined by
\begin{equation}
\chi^2 = \sum_{i=1}^N\sum_{j=1}^N (D_i-T_i(a))C^{-1}_{ij}
(D_j-T_j(a)),
\label{l}
\end{equation}
where $N$ is the number of data points, $D_i$ is the 
measurement
and $T_i(a)$ is the theoretical prediction depending on parton input 
parameters $a$. Alternatively, one can incorporate 
the systematic errors into the theory prediction
\begin{equation}
f_i(a,s) = T_i(a) + \sum_{k=1}^n s_k \Delta_{ik},
\label{m}
\end{equation}
where $\Delta_{ik}$ is the one-sigma correlated error for point 
$i$ from source $k$. The $\chi^2$ is then defined by     
\begin{equation}
\chi^2 = \sum_{i=1}^N \biggl(\frac{D_i-f_i(a,s)}{\sigma_{i,unc}}
\biggr)^2 + \sum_{k=1}^n s_k^2,
\label{n}
\end{equation}
where the second term constrains the values of $s_k$, assuming that 
the correlated systematic errors are Gaussian distributed.
This is identical to the correlation matrix 
definition of $\chi^2$ if the errors are small. 
In many cases the statistical and systematic errors are simply added in 
quadrature, either for simplicity, because this has much the same result as 
the above procedures, or due to lack of information on the correlations 
of systematic errors. (The sophisticated treatment is found to be 
essential for Tevatron jets where correlated systematic errors dominate.)   

All groups performing global fits use some combination of the above ways to
calculate $\chi^2$, and minimizing with respect to the defined $\chi^2$ 
completely determines the parton distributions. In general, the quality of 
the total fit is reasonably good, e.g. for CTEQ6 one obtains 
$\chi^2/d.o.f.=1954/1811$ and for MRST2001 $\chi^2/d.o.f.=2328/2097$.

\vspace{-0.5cm}
\begin{figure}[htbp]
\begin{center}
\includegraphics[width=9cm]{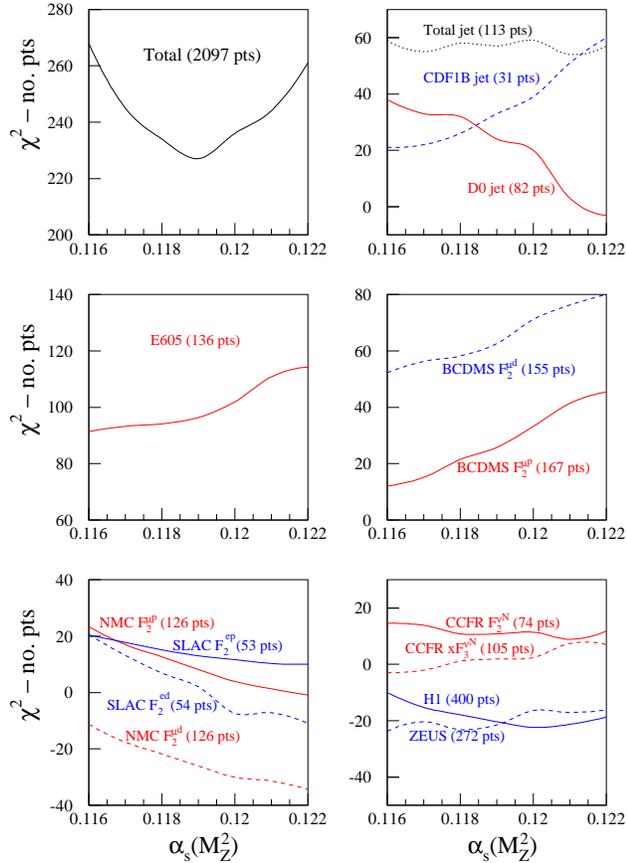}
\end{center}
\vspace{-2.5cm}
\caption{The quality of the fit to the individual data sets
included in the global analysis by MRST, shown together with the grand
total $\chi^2$, as a function of $\alpha_S(M_Z^2)$}
\label{one}
\end{figure}

In principle, the one-$\sigma$ error on some parameter in the fit, e.g. the 
value of $\alpha_S(M_Z^2)$ or one of the parameters describing the input 
parton distributions, can be determined by allowing the value of $\chi^2$ 
to vary one unit from its minimum. In practice, this results in unrealistically
small errors. This is shown in Fig.~1, which shows the variation of the 
$\chi^2$ with $\alpha_S(M_Z^2)$ for the total fit, and for individual sets 
within the global fit for the MRST partons.    

It is also demonstrated in an alternative manner in Fig.~2, where for the 
CTEQ6 fit
the extraction of $\alpha_S(M_Z^2)$ obtained by allowing $\chi^2$ to increase
by one unit for each data set within the fit is presented
\cite{wuki}. Clearly the size of 
the errors and the scatter of the points are unrealistic. 

\vspace{2.6cm}

\begin{figure}[htbp]
\begin{center}
\includegraphics[width=7.5cm]{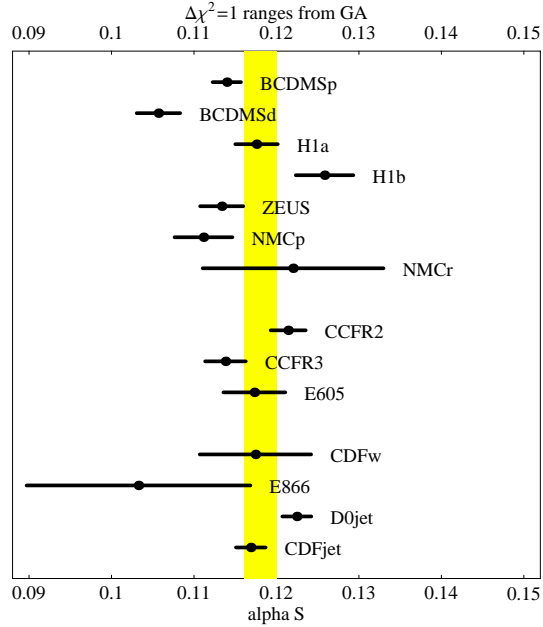}
\end{center}
\vspace{-4cm}
\caption{Values of $\alpha_S(M_Z^2)$ obtained by CTEQ from the quality of 
fit to individual data sets within the global fit.}
\label{two}
\end{figure}

This  failure of the $\Delta \chi^2=1$ rule may have various sources. 
Firstly, it is clear from examination that not all of the data sets are 
consistent with each other, i.e. the errors have been underestimated in 
some manner. Thus fitting the data simultaneously leads to a certain degree 
of inconsistency. One could keep only those sets which are clearly 
consistent with each other (within the context of NLO QCD) 
\cite{ref:giele}. However, this leads
to a severe restriction in the amount of data used, and 
the partons in some regions would lose important constraints.  
(Due to the correlation between different regions of $x$ via sum rules
and because evolution equations for partons involve convolutions over
a range of $x$, problems in one range of $x$ impact on the partons over 
the whole range.) 
A second problem is that certain assumptions have to be made when 
performing a global fit. These include the cuts made on data, 
data sets included in the fit, the parameterization for input parton sets, 
the form of strange sea, the precise heavy flavour 
prescription,  the definition of the NLO strong coupling constant, 
the starting scale of evolution {\it etc.} Many of these can lead to 
variations considerably greater 
than those obtained from $\Delta \chi^2=1$.     
Finally, we do not have a perfect theory -- NLO-in-$\alpha_S$ QCD has many 
corrections. Obviously there are higher orders (NNLO), which are not that 
small in themselves in QCD, but there are other problems. In 
particular at large and small $x$ there are large logarithms associated 
with higher orders, i.e.  
at small $x$ there are terms of the type $\alpha_s^n \ln^{n-1}(1/x)$, 
and at large $x$ terms like $\alpha_s^n 
\ln^{2n-1}(1-x)$. These are not fully understood.  Also at low $Q^2$ 
there are higher twist, i.e. ${\cal O}(1/Q^2)$ corrections that are again not 
well understood. Hence, in fitting data one parameter may be artificially 
well constrained at some incorrect value in some region of parameter space 
by the necessity to account for missing corrections to the theory. This can 
then influence other regions in the manner explained above, e.g. an 
artificially large gluon in one range of $x$ can lead to it being too small
in other regions because of the momentum sum rule constraint.

This has led to various methods to determine errors by allowing a wider 
variation than the canonical $\Delta \chi^2=1$. CTEQ actually allow 
$\Delta \chi^2=100$ by various considerations, including error limits of 
individual experiments \cite{ref:lmethod}. 
An example is the determination of 
$\alpha_S(M_Z^2)$ as shown in Fig.~3, where the error bars give the $90\%$
confidence limit for each experiment (the size of which varies considerably 
between data sets), and the two lines give the band of 
$\alpha_S(M_Z^2)$ allowed by $\Delta \chi^2=100$ for the global fit 
\cite{wuki}. Arguably this is too conservative, and CTEQ regard 
it as more than a one-$\sigma$ error. 
MRST use $\Delta \chi^2=20$ as an approximate one-$\sigma$ 
error instead, based on 
the essentially subjective criterion of judging when the fit is
going uncomfortably wrong in some region \cite{MRST01}.

\vspace{2.8cm}

\begin{figure}[htbp]
\begin{center}
\includegraphics[width=7.5cm]{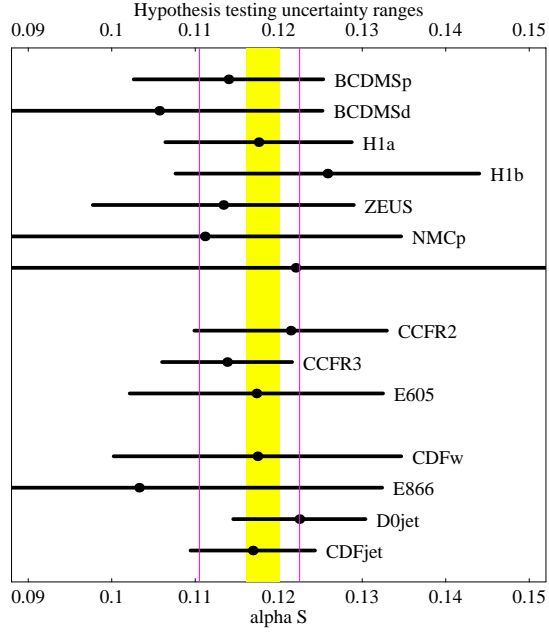}
\end{center}
\vspace{-4cm}
\caption{Allowed value of $\alpha_S(M_Z^2)$ obtained by CTEQ from the 
quality of fit to individual data sets within global fit.}
\label{three}
\end{figure}

A more precisely defined method is to use 
an alternative formulation for $\chi^2$. 
In the offset method the fit is obtained by minimizing
\begin{equation}
\chi^2 = \sum_{i=1}^N \biggl(\frac{(D_i-f_i(a,s))}{\sigma_{i,unc}}
\biggr)^2,
\label{w}
\end{equation}
i.e. the best fit and parameters $a_0$ are obtained from only 
uncorrelated errors. 
The quality of the fit can then be
estimated by adding errors in quadrature. The systematic errors on 
$a_i$ are determined by letting each $s_k = \pm 1$ and adding the 
deviations in quadrature. This is essentially the same as, and in practice 
achieved by calculating 2 
Hessian matrices
\begin{equation}
M_{ij} = \frac {\partial ^2 \chi^2}{\partial a_i \partial a_j}
\qquad  V_{ij} = \frac {\partial ^2 \chi^2}{\partial a_i \partial s_j},
\label{x}
\end{equation}
and defining covariance matrices  
\begin{equation}
C_{stat} = M^{-1} \qquad C_{sys} = M^{-1}VV^TM^{-1} \qquad 
C_{tot} = C_{stat} + C_{sys}.
\label{y}
\end{equation}
This method was used in early H1 fits \cite{ref:pascaud} and in 
ZEUS fits \cite{ref:mbfit,ref:zeusfit}. Much the same method is used for 
polarized data in \cite{spin}, as far as the information on correlated 
errors allows. Strictly speaking, the method is not optimum, and it 
leads to a much bigger uncertainty than the standard method  
for $\Delta \chi^2=1$ \cite{ref:alekhinstudy}. 
However, it may be suitable in practice \cite{ref:zeusfit}, and 
ZEUS estimate the same results could be obtained 
using the standard method and $\Delta \chi^2=50$. 

The results of the various approaches may be summarized in a selection of 
the extracted values of $\alpha_S(M_Z^2)$ below \cite{vladimir}. 
\begin{eqnarray}
\hbox{CTEQ6} \quad (\Delta \chi^2 = 100) \quad  \alpha_S(M_Z^2) &=& 0.1165 \pm
0.0065 \nonumber \\ 
\hbox{ZEUS} \quad (\Delta \chi^2_{eff} = 50) 
\quad  \alpha_S(M_Z^2) &=& 0.1166 \pm
0.0008 (\hbox{uncor}) \pm 0.0032(\hbox{corr}) \nonumber\\
& &\pm 0.0036(\hbox{norm})\pm 0.0018(\hbox{model}) \nonumber \\   
\hbox{MRST}  \quad(\Delta \chi^2 = 20) \quad  \alpha_S(M_Z^2) &=& 0.119 \pm
0.002 (\hbox{exp}) \pm 0.003(\hbox{theory}) \nonumber \\
\hbox{H1} \quad (\Delta \chi^2 = 1) \quad  \alpha_S(M_Z^2) &=& 0.115 \pm
0.0017 (\hbox{exp}) \pm 0.005(\hbox{theory})
\label{alphas}
\end{eqnarray}

The values obtained are consistent, and the errors not too
dissimilar given the wide variation in $\Delta \chi^2$ used. This is largely 
because each group has chosen a method which gives a reasonable and believable
error. H1 actually use the standard definition, obtaining a small
error, but are only able to do this because they use a much smaller data 
set than the other groups, i.e. their own data plus some BCDMS data which 
are fully consistent. If MRST and CTEQ were to use the same approach their
error would be tiny (and their extracted value of $\alpha_S(M_Z^2)$
some standard deviations away from that of H1, e.g. MRST2001 would obtain 
$\alpha_S(M_Z^2) = 0.1192 \pm 0.0005$).   

\section{Conclusions}

In determining errors for and from parton distributions one encounters a 
dilemma. One could fit to a relatively small subset of the available data 
with a very conservative set of cuts so that data are self consistent and
theory is truly reliable. However, this would lead to a poor determination 
of partons in certain regions and would give no idea how well the theory is 
working overall. Alternatively, one can try to constrain the partons in as wide
a region of parameter space as possible, accepting that the standard 
rules for error determination will break down. The way in which 
one can obtain sensible estimates of the real errors in the latter
case is a subject very much open for discussion.      

\subsection*{Acknowledgements}

We would like to thank some of our colleagues not at the discussion,    
A.D. Martin, J. Pumplin, R.G. Roberts and W.K. Tung for useful comments 
and the workshop organizers L.Lyons and M. Whalley.

\end{document}